\documentclass[twocolumn]{aastex631}

\usepackage{float,graphicx,amsmath,booktabs,natbib,tabularx,multirow,hyperref}
\usepackage[caption=false]{subfig}

\newcommand{\kms}{km\,s$^{-1}$\ }

\definecolor{dark-red}{rgb}{0.4,0.15,0.15}
\definecolor{dark-blue}{rgb}{0.15,0.15,0.4}
\definecolor{medium-blue}{rgb}{0,0,0.5}
\hypersetup{
    colorlinks, linkcolor={dark-red},
    citecolor={dark-blue}, urlcolor={medium-blue}
}

\shortauthors{Sharma et al.}

\graphicspath{{./}{figures/}}

\def \caltech {{Division of Physics, Mathematics and Astronomy, 
California Institute of Technology, Pasadena, CA 91125, USA}}

\def \coo {{Caltech Optical Observatories, California Institute of Technology, Pasadena, CA 91125, USA}}

\def \su {{Department of Astronomy, The Oskar Klein Center, Stockholm University, AlbaNova, 10691 Stockholm, Sweden}}

\begin{document}

\title{CCSNscore: A multi-input deep learning tool for classification of core-collapse supernovae using SED-Machine spectra}

\correspondingauthor{Yashvi Sharma}
\email{yssharma@astrocaltech.edu}

\author[0000-0003-4531-1745]{Yashvi Sharma}
\affiliation{\caltech}

\author[0000-0003-2242-0244]{Ashish A. Mahabal}
\affiliation{\caltech}
\affiliation{Center for Data Driven Discovery, California Institute of Technology, Pasadena, CA 91125, USA}

\author[0000-0003-1546-6615]{Jesper Sollerman}
\affiliation{\su} 

\author[0000-0002-4223-103X]{Christoffer Fremling}
\affiliation{\coo}

\author[0000-0001-5390-8563]{S. R. Kulkarni}
\affiliation{\caltech}

\author[0000-0002-5683-2389]{Nabeel Rehemtulla}
\affiliation{Department of Physics and Astronomy, Northwestern University, 2145 Sheridan Road, Evanston, IL 60208, USA}
\affiliation{Center for Interdisciplinary Exploration and Research in Astrophysics (CIERA), Northwestern University, 1800 Sherman Ave, Evanston, IL 60201, USA}

\author[0000-0001-9515-478X]{Adam~A.~Miller}
\affiliation{Department of Physics and Astronomy, Northwestern University, 2145 Sheridan Rd, Evanston, IL 60208, USA}
\affiliation{Center for Interdisciplinary Exploration and Research in Astrophysics (CIERA), Northwestern University, 1800 Sherman Ave, Evanston, IL 60201, USA}

\author{Marie Aubert}
\affiliation{Universit\'e Clermont Auvergne, CNRS/IN2P3, LPCA, F-63000 Clermont-Ferrand, France}

\author[0000-0001-9152-6224]{Tracy X. Chen}
\affiliation{IPAC, California Institute of Technology, 1200 E. California Blvd, Pasadena, CA 91125, USA}

\author[0000-0002-8262-2924]{Michael W. Coughlin}
\affiliation{School of Physics and Astronomy, University of Minnesota, Minneapolis, Minnesota 55455, USA}

\author[0000-0002-3168-0139]{Matthew J. Graham}
\affiliation{\caltech}

\author{David Hale}
\affiliation{\coo}

\author[0000-0002-5619-4938]{Mansi M. Kasliwal}
\affiliation{\caltech}

\author[0000-0002-1031-0796]{Young-Lo Kim}
\affiliation{Department of Physics, Lancaster University, Lancs LA1 4YB, UK}

\author[0000-0002-0466-1119]{James D. Neill}
\affiliation{\coo}

\author[0000-0003-1227-3738]{Josiah N. Purdum}
\affiliation{\coo}

\author[0000-0001-7648-4142]{Ben Rusholme}
\affiliation{IPAC, California Institute of Technology, 1200 E. California Blvd, Pasadena, CA 91125, USA}

\author[0000-0003-2091-622X]{Avinash Singh}
\affiliation{\su}

\author{Niharika Sravan}
\affiliation{Department of Physics, Drexel University, Philadelphia, PA 19104, USA}

\begin{abstract}
Supernovae (SNe) come in various flavors and are classified into different types based on emission and absorption lines in their spectra. SN candidates are now abundant with the advent of large systematic sky surveys like the Zwicky Transient Facility (ZTF), however, the identification bottleneck lies in their spectroscopic confirmation and classification. Fully robotic telescopes with dedicated spectrographs optimized for SN follow-up have eased the burden of data acquisition. However, the task of classifying the spectra still largely rests with the astronomers. Automating this classification step reduces human effort and can make the SN type available sooner to the public. For this purpose, we have developed a deep-learning based program for classifying core-collapse supernovae (CCSNe) with ultra-low resolution spectra from the SED-Machine spectrograph on the Palomar 60-inch telescope. The program consists of hierarchical classification task layers, with each layer composed of multiple binary classifiers running in parallel to produce a reliable classification. The binary classifiers utilize RNN and CNN architecture and are designed to take multiple inputs to supplement spectra with $g$- and $r$-band photometry from ZTF. On non-host-contaminated and good quality SEDM spectra (``gold" test set) of CCSNe, CCSNscore is $\sim$94\% accurate in distinguishing between hydrogen-rich (Type~II) and hydrogen-poor (Type~Ibc) CCSNe. With light curve input, CCSNscore classifies $\sim83$\% of the gold set with high confidence (score $\geq 0.8$ and score-error $<0.05$), with $\sim98$\% accuracy. Based on SNIascore's and CCSNscore's real-time performance on bright transients ($m_{pk}\leq18.5$) and our reporting criteria, we expect $\sim0.5\%$ ($\sim4$) true SNe~Ia to be misclassified as SNe~Ibc and $\sim6\%$ ($\sim17$) of true CCSNe to be misclassified between Type~II and Type~Ibc annually on the Transient Name Server.

\end{abstract}

\section{Introduction} \label{sec:intro}
Wide-field optical transient surveys are already finding supernova (SN) candidates in record numbers (\citealp[ASAS-SN]{asassn}; \citealp[PS1]{chambers2016panstarrs1}; \citealp[ATLAS]{atlas}; \citealp[ZTF]{Bellm2019b,graham2019,Dekany20}) which will increase by tenfold in the era of the Rubin Observatory \citep{lsst}. A supernova candidate becoming a secure SN identification involves many steps. Taking the example of ZTF, the transient `alerts' from ZTF \citep{Patterson2019} are filtered by alert management frameworks (e.g.\ \texttt{Fritz} \citep{skyportal2019, Coughlin2023}, AMPEL \citep{ampel}, etc.) to obtain potential SN candidates from the slurry of transients, variable stars, moving solar system objects, and bogus artifacts. These candidates are visually inspected for spectroscopic follow-up candidates; however, this step can be automated depending on survey needs as demonstrated in \citet{Rehemtulla2024} (\texttt{BTSbot}). Finally, the selected candidates are assigned to various telescope facilities to obtain secure spectroscopic classifications, but as follow-up resources are limited, this step becomes the primary bottleneck. Still, dedicated SN classification instruments and programs (e.g., \citealp[SEDM]{Ben-Ami12,sedm2018}; \citealp[ePESSTO]{pessto}; \citealp[The Global Supernova Project]{GSNP}) take spectra of a few thousand SNe per year, which are then analyzed by astronomers, assigned a classification and then sent to the Transient Name Server (TNS\footnote{\url{https://www.wis-tns.org/}}). Some programs exist that are meant to support astronomers in the manual SN classification task, such as  SuperNova IDentification software (SNID; \citealp{SNID2007}), \texttt{Superfit} \citep{superfit}, NGSF \citep{ngsf}, Gelato \citep{gelato2008}, all based on either template cross-correlation techniques \citep{TonryDavis1979} or minimization algorithms. These programs often require user input (initial guesses for redshift, age, restriction of parameter search ranges, etc.) to obtain correct classifications. Still, they can be less effective because of host contamination (in SNID) or poor signal-to-noise ratio (SNR). Moreover, template-matching techniques are slow to run on thousands of spectra, suffer from type-attractor issues if one kind of template dominates the template bank, and are less accurate when automated \citep{kim_2024}. With the advent of deep learning techniques and the dedicated influx of spectral data, sophisticated deep learning-based models can be trained to automatically and reliably classify the most common SN types. \citet{Muthukrishna_2019} presented DASH (Deep Learning for the Automated Spectral Classification of Supernovae and their Hosts), trained on SNID template dataset (which contains intermediate resolution spectra) and tested on the OzDES \citep{ozdes2015a,ozdes2017} dataset, also from intermediate resolution (R $\sim1400$) spectrographs. Though DASH showed promising performance on the OzDES test set and is easy to install and use, it did not perform well on ultra-low resolution spectra (R $\sim100$) when tested in \citet{Fremling_2021,kim_2024}. 

Thus, \textit{SNIascore} \citep{Fremling_2021} -- a deep-learning based binary classifier was developed specifically for classifying SNe~Ia using the spectra taken by SED-Machine \citep{Ben-Ami12,sedm2018,pysedm2019,Kim2022}, an ultra-low resolution (R$\sim100$) IFU spectrograph operating in the optical wavelength range (3800\,\AA\ -- 9150\,\AA) on the fully robotic Palomar 60-in telescope (P60; \citealp{Cenko2006}). The need for \textit{SNIascore} was motivated by the ZTF Bright Transient Survey (BTS; \citealp{Fremling2020,Perley2020,Rehemtulla2024}); a flux-limited survey to spectroscopically classify bright transients ($m_{peak}<18.5$) detected by ZTF with $>90$\% completeness. With SEDM's resolution on the moderate aperture of P60, it is uniquely suited for bright transient classification and thus became the main workhorse instrument for BTS, as well as the top classifier on TNS. \textit{SNIascore} was optimized to classify SNe~Ia with more than 90\% accuracy at less than 0.6\% false positive rate (FPR), and with this performance automated half of the manual classification workload for BTS. \textit{SNIascore} has allowed BTS to send robust SN~Ia classifications to the TNS within $\sim11$ minutes of acquisition. Together, \texttt{BTSbot} and SNIascore enabled the first fully automatic end-to-end discovery and classification of an optical transient \citep{Rehemtulla_TNS}.

As (normal) SNe~Ia are quite homogeneous in their spectral and photometric properties, it is a binary classification problem suited for deep learning. Also, SNe~Ia are the most abundant type of supernova identified in flux-limited surveys like BTS, thus providing a large sample sufficient for training deep learning models. Core-collapse (CC) supernovae, however, are more heterogeneous, with some CCSNe even transitioning to a different spectral type over time or developing late-time interaction signatures \citep{sn2014c,P1_2017ens,sollerman2020,Sharma2024,Kangas2024}. In BTS \citep{Perley2020}, the most abundant class among CCSNe is the hydrogen-rich Type~II SNe ($\sim72\%$) followed by the hydrogen-poor Type~I or stripped-envelope SNe (SESNe). Within the hydrogen-rich Type~II class, the most common ($\sim76\%$) are the spectroscopically ``normal" subtypes (SNe~IIP/L) showing strong P-Cygni Balmer line profiles, with the rest (SNe~IIb, IIn, and SLSN-II) contributing $\sim24\%$ combined. Within the Type~I SESNe, SNe~Ibc consitute $\sim59\%$, SNe~Ic-BL make up $\sim19\%$, SLSN-I are $\sim13\%$ and the rest are the rare SNe~Ibn and SNe~Icn (SN subtype fractions referenced from \citealp{Perley2020}). This is a highly unbalanced dataset with the rarer subtypes only having a handful of examples, not nearly enough for training a deep-learning model. The problem is compounded by varying levels of noise in the spectra and the classification being inherently difficult for some subtypes with ultra-low resolution spectra (for example, the `n' in IIn and Ibn refers to `narrow' spectral lines of $\sim100$\,\kms, impossible to resolve with a resolution of R $\sim100$). For these reasons, developing a high-performing and reliable automated spectral classifier for CCSNe using just the SEDM data is challenging.

This work attempts to face this challenge and presents a deep learning-based program -- CCSNscore, designed specifically for CCSN classification, trained with SEDM spectral data, Open SN Catalog \citep{opensncatalog2017} spectral data, and ZTF photometry. The data preparation and preprocessing are described in \S\ref{sec:data}, the application structure and model architecture are described in \S\ref{sec:model}, the training and optimization process is outlined in \S\ref{sec:training} and the performance on the test set is detailed in \S\ref{sec:results}. We explore the limitations and caveats of this tool in \S\ref{sec:disc}. The CCSNscore software, trained models presented in this paper, and the metadata of training and test datasets are available on the GitHub repository of \href{https://github.com/Yashvi-Sharma/CCSNscore}{CCSNscore}.

\section{Dataset}\label{sec:data}

\begin{figure*}[thbp!]
\centering
    \includegraphics[width=0.98\textwidth]{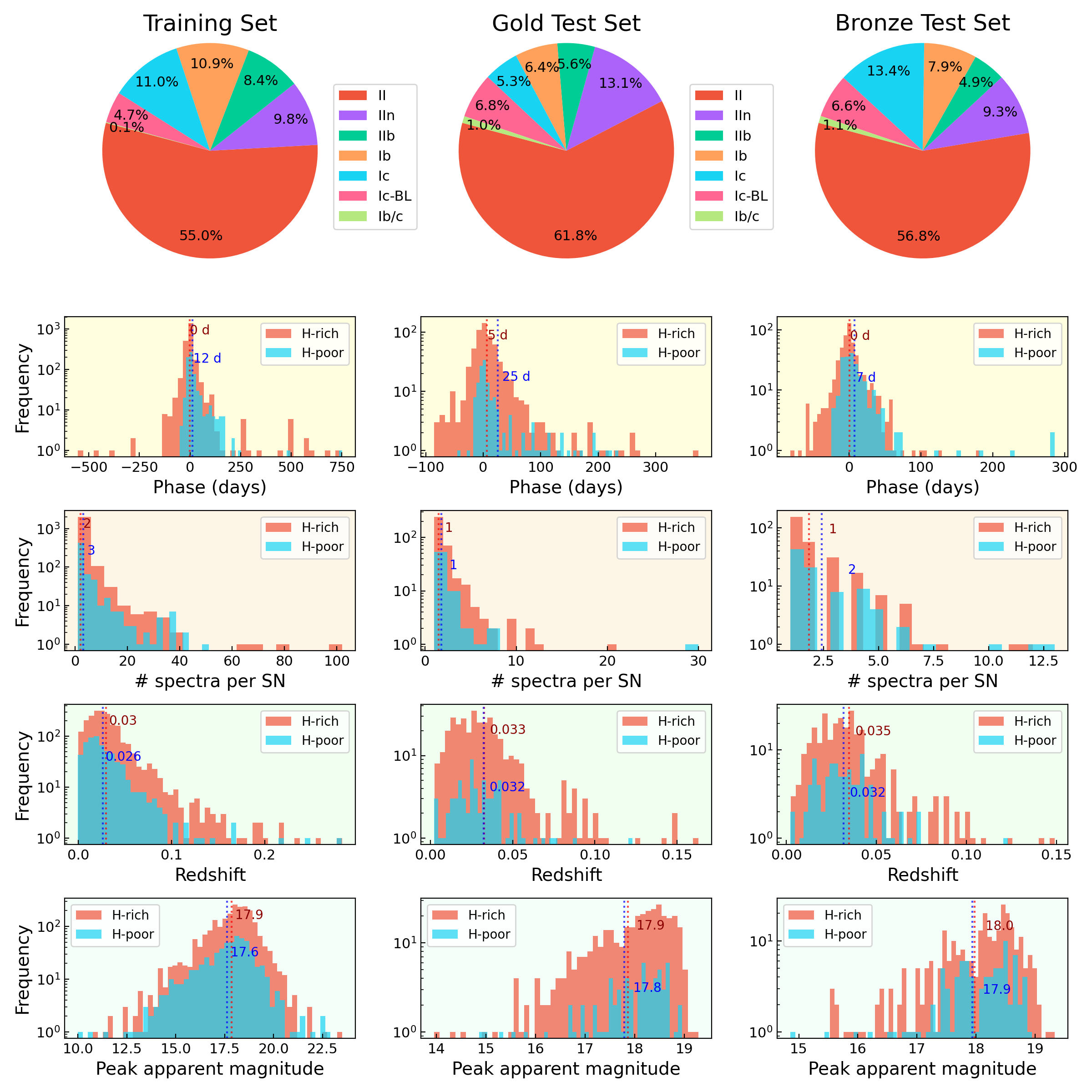}
    \caption{Distribution of properties of the training and test sets (gold and bronze). The first row contains pie charts depicting the highly imbalanced distribution of samples by CCSN subtype. The next four rows depict the property distribution for hydrogen-rich (Type~II in blue) and hydrogen-poor (Type~Ibc in red) SN samples separately, with the second row showing spectral phase distribution, the third row showing the distribution of the number of spectra per unique SN, the fourth row showing redshift distribution of the unique SNe and the fifth row showing the peak apparent magnitude distribution of unique SNe. The mean values of all distributions are marked with dotted vertical lines.}
    \label{fig:typedist}
\end{figure*}

Initially, we started our study with the SEDM spectra of BTS transients used in \citet{Fremling_2021} for \texttt{SNIascore} that are not SNe~Ia (`non-Ia') and collected between March 2018 and August 2020. We kept the BTS sample spectra published in \citet{Fremling2020} in the test set along with the spectra of a few peculiar CCSNe and put the rest in the training set. We made sure that all spectra belonging to a transient were present in only one of the sets. We updated the dataset a few times as more spectra were collected through BTS, until March 2024. We also added optical spectra from the Open SN Catalog and resampled them to match the resolution of the SEDM spectra. Because we wanted to test the performance primarily on SEDM data, we put all of the Open SN Catalog spectra of SNe that had an unambiguous and non-peculiar classification in the training set. Then, we split the newer SEDM spectra of ZTF transients into training and test sets such that the final ratio of test samples to training samples per major subtype would be between 10\% to 30\%. We put most of the stripped-envelope SNe with ambiguous subtype (Type~Ib/c) in the test set. 

There are 8563 unique spectra in our training set, of which 3015 are SEDM spectra of 1222 unique ZTF transients and 5548 are Open SN Catalog spectra of 1546 unique transients. The training data consists of a broad range of spectral quality, from poor to great SNR (excluding extremely noisy cases), various levels of host galaxy contamination, and various strengths of the emission and absorption features (including completely featureless spectra), so that the models can learn to expect all kinds of observed data and do not overfit on only good quality of data. The test data containing 1535 SEDM spectra also shows this wide variety of spectral quality. To assess the model performance on good vs. bad (unclassifiable or difficult to classify) quality spectra, we split the test data into ``gold" and ``bronze" categories semi-automatically through a combination of parameter thresholding and visual inspection. The classifiability of a spectrum depends not just on the noise and SNR but also on the presence of broad supernova features (or similarity to SNe), which SNID encapsulates well. We found that the number of ``good" SNID matches with an \footnote{$rlap$ is a SNID parameter indicating the goodness of a template fit}$rlap > 4$ (referred to as $numSNID$ henceforth) serves well as a discriminator for gold vs. bronze split. We found $numSNID \geq 20$ appropriate for crudely separating spectra with clear supernova features and decent SNR. Next, we visually inspected all Type~Ibc (hydrogen-poor) spectra in the test sample to identify spectra with severe host contamination. If the host galaxy has strong typical narrow emission lines (H$\alpha$, \ion{O}{3}), they show up as blended emission lines in SEDM spectra due to the ultra-low resolution and appear just like the features of a SN~II or SN~IIn in SEDM spectra, making automatic identification of the host-contaminated Type~Ibc SNe extremely difficult. We keep all of the visually identified host-contaminated samples in the bronze set. Finally, the gold (bronze) test set has a total of 780 (755) SEDM spectra of 431 (369) unique ZTF transients. The split comes out to be nearly 50\%.

The properties of the (non-augmented) training and test set samples are shown in  Figure~\ref{fig:typedist}. From the subtype distribution plots in Figure~\ref{fig:typedist} (top row), it is immediately obvious that there is a significant imbalance between the subsamples of the hydrogen-rich (Type~II) and the hydrogen-poor (Type~Ibc) SNe. To correct this imbalance, we augment the data with fake spectra for each subtype by randomly choosing pairs of samples (of that subtype) around similar ages, transforming their wavelengths to rest-frame (deredshifting), and taking their weighted average (random weights). The number of fake samples to create per subtype can be chosen during the data preprocessing step. The fake samples are added to the real data to balance the distribution of training samples across categories for a given classification task. The second row of Figure~\ref{fig:typedist} shows the spectral phase (days from maximum brightness at the time of spectral observation) distribution. The time of maximum brightness is not always equal to the peak brightness of the SN as photometric coverage is sparse for many SNe in the dataset, often missing the rise and the peak. The mean (sigma-clipped) of the H-rich SNe phase distribution is around the maximum brightness while it is after the maximum brightness for the H-poor SNe phase distribution for all three sets (training, gold test, bronze test). The standard deviation of the H-rich distribution is also higher than the H-poor distribution for all three sets, which reflects the naturally longer duration of H-rich SNe resulting in more follow-up observations. The third row of Figure~\ref{fig:typedist} depicts the distribution of the number of spectra per unique supernova, the mean of which is centered around 2--3 spectra per SN. The training set has more SNe with thorough spectral series data ($>10$ spectra) owing to the dataset from the Open SN Catalog. The test sets, being exclusively SEDM only have $<10$ SNe per set that have $\geq10$ spectra, not significant to affect the performance. These multiple spectra also probe various levels of noise, effects of varying sky background, and age of the SNe and thus in a way are unique to the models. The fourth and fifth rows show the redshift and peak apparent magnitude distributions of the unique SNe. The redshifts are of the host-galaxy of the SNe when available through the NED database or derived from the SNe classification spectra. The peak apparent magnitude distribution shown in the last row of Figure~\ref{fig:typedist} is highly dependent on the photometric coverage quality as mentioned earlier. The peak magnitudes were obtained from the BTS sample explorer page when available or derived from the interpolated light curves. The distributions center around 18 mag, and for the test data drop sharply around 19 mag as that is the maximum depth attainable by SEDM for reasonable exposure times.

\begin{figure*}[htbp!]
    \centering
    \includegraphics[width=0.95\textwidth]{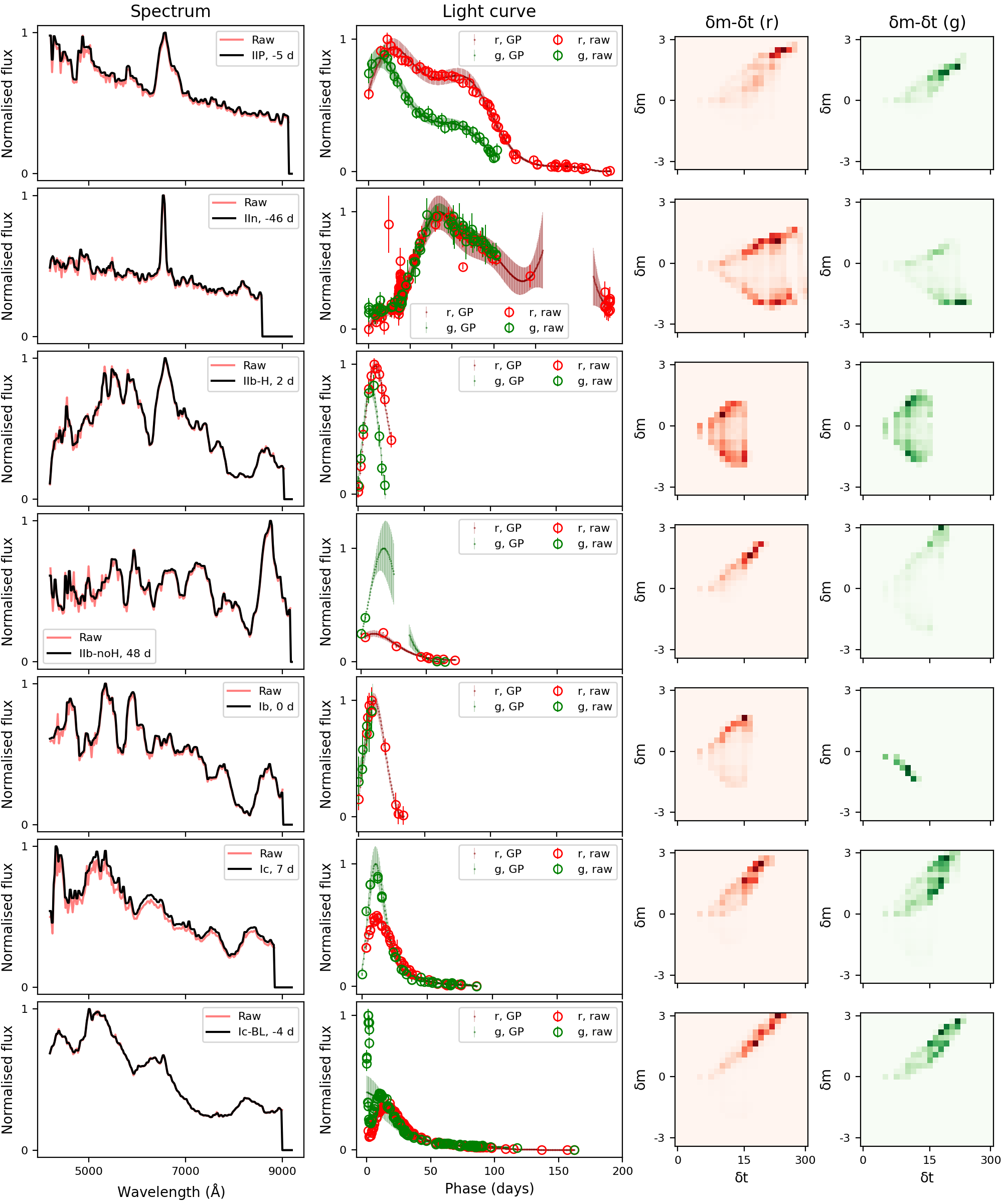}
    \caption{Training data samples representing the various transient types (rows). The first column shows the raw (red) and processed (black) normalized spectra, the second column shows the raw and GP-interpolated $r$- and $g$-band normalized light curves, and the third and fourth columns show $\delta m-\delta t$ representations of the $r$- and $g$-band GP-interpolated light curves, respectively.}
    \label{fig:data}
\end{figure*}

To further help with the classification task, we added the capability to use ZTF $g$- and $r$-band light curves \citep{Masci2019} as additional input channels. The light curves can be supplied as fixed-length flux arrays where the fluxes are taken from the first detection of the transient in ZTF to a set number of days (by default 200 days past first detection). The fixed length of arrays is a requirement of the model architecture. Both the SEDM spectra and ZTF light curves were queried from the dynamic user interfaces, the GROWTH Marshal \citep{Kasliwal_2019} and \texttt{Fritz} \citep{skyportal2019,Coughlin2023}. No light curve information is currently supplied with the Open SN catalog spectra. We also added another method of providing the light curves by transforming them into ``$\delta m-\delta t$" phase space \citep{Mahabal_2017}. The $\delta m-\delta t$ representation takes all pairs of light curve points and maps them to a 2D space with the y-axis being the magnitude difference and the x-axis being the time difference (in days) between the two points. This 2D space can potentially capture the different rise and decline rates of various subtypes. The program offers optional usage of these additional channels for training. Further details on data preprocessing are described in the sections below. Figure~\ref{fig:data} shows examples of input data samples for the various subtypes. A data ``sample" in this study refers to one spectrum with its corresponding $g,r$ light curves, and their $\delta m-\delta t$ representations. Note that there can be multiple spectra of the same supernova but each spectrum is counted as an individual sample. We have split the Type IIb class into `IIb-H' (spectra at phases in which H$\alpha$ P-Cygni dominates) and `IIb-noH' (spectra at phases when the H$\alpha$ feature has weakened and turned towards the nebular phase) to put in the H-rich and H-poor classes. Except the IIn and IIb-noH examples all shown examples are representative of spectra taken near peak-light and are good quality, not host-contaminated SEDM spectra that show strong and clear SN features. However, as mentioned earlier, the training samples span a wide range of SNR and SN feature strengths.

\subsection{Preprocessing}
\subsubsection{Optical spectra}
All spectra are deredshifted using the redshifts obtained from the GROWTH Marshal and \texttt{Fritz} for ZTF transients and the Open SN Catalog for the rest. Though SEDM spectra cover the wavelength range between 3700 \AA\ and 9200 \AA\, the bluest part of the spectra have been observed to be noisy. Therefore, we restrict the deredshifted wavelength range from 4200 \AA\ to 9200 \AA\ and interpolate the fluxes with a cubic spline function to get fluxes at a fixed space wavelength array of 256 points (to match the sampling across SEDM and Open SN catalog spectra) in the mentioned wavelength range. This flux array is then median filtered, and normalized, and any `nan' values are converted to zero. We do not divide the spectra by the continuum as the continuum also contains information relevant to supernova classification.

\subsubsection{1D light curves}
We take the 5$\sigma$ detections from the ZTF $g$- and $r$-band light curves and fit them using Gaussian process (GP) regression with brightness as the dependent variable (in magnitudes) and phase from first detection as the independent variable (in days). Interpolation is necessary as the cadence of the light curves is not constant and there are gaps in the data due to weather, sun occultation, and instrument downtime. We use a combination of a radial basis function (RBF, also known as ``squared-exponential") kernel with length scale bounds between (20,200) days, and a White Kernel to characterize the noise. We interpolate magnitudes and magnitude errors between the first and last detection in 1~day bins. Since all the input samples to the model need to be of fixed length, we pad the interpolated light curves with zeros if they are shorter than 200~days, and truncate if they are longer. These fixed length magnitudes and magnitude error arrays are then converted to linear fluxes (erg\,s$^{-1}$\,cm$^{-2}$\,\AA$^{-1}$) and normalized. The light curves are not redshifted or K-corrected to limit preprocessing steps, especially the ones that rely on prior redshift being available. The final input that goes to the model is a stacked array with fluxes at the $0^{th}$ index and flux errors at the $1^{st}$ index, with a shape of (200,2) per sample.

\subsubsection{$\delta m-\delta t$ light curves}
From the interpolated light curves, we compute the magnitude and time (in days) difference for all pairs of light curve points and create a 2D histogram with fixed but non-uniform bin sizes in magnitude and phase space. The bin intervals are decided based on the magnitude ranges that our transients span and their typical duration timescales. The magnitude bin edges are [$-4.5$, $-3$, $-2.5$, $-2$, $-1.5$, $-1.25$, $-0.75$, $-0.5$, $-0.3$, $-0.2$, $-0.1$, $-0.05$, 0, 0.05, 0.1, 0.2, 0.3, 0.5, 0.75, 1.25, 1.5, 2, 2.5, 3, 4.5] mags, and the phase bin edges are [0, $\frac{1}{24}$, $\frac{4}{24}$, 0.25, 0.5, 0.75, 1, 2, 3, 6, 9, 12, 15, 24, 33, 48, 63, 72, 96, 126, 153, 180, 216, 255, 300] days. The shape of these 2D histogram inputs is (24,24) per sample. Examples of $\delta m-\delta t$ histograms for the SN subtypes are shown in Figure~\ref{fig:data}.

\subsubsection{Simulated data for augmentation}
The number of synthetic samples to create per subtype is specified during the preprocessing step. For a given subtype, we create two subsets from its training samples depending on the phase of the spectrum, the first subset (early-time phase) contains spectra taken before the light curve maximum, and the second subset (photospheric phase) has spectra taken after the light curve maximum till the SN becomes nebular. We use the early-time subset to create 30\% of the total synthetic samples and the photospheric subset for the remaining 70\%. To make a synthetic sample, we randomly choose pairs of samples from a subset without replacements and take the weighted average of their rest-frame (deredshifted) spectra with randomly chosen weights to generate the synthetic spectrum. We also set the synthetic sample's redshift to the weighted average of the pair's redshifts. Then, we pick one of the light curves from the pair, scale its flux to the new luminosity distance (redshift), and use it to generate the synthetic 1D interpolated light curves and their $\delta m-\delta t$ representations for the synthetic sample.

\section{Parallel binary classifiers}\label{sec:model}

\begin{figure*}
    \centering
    \includegraphics[width=0.9\textwidth]{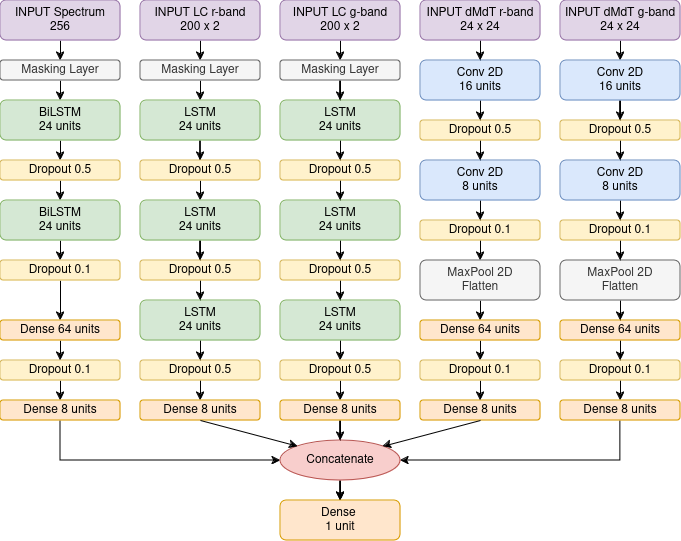}
    \caption{Multi-channel network architecture of CCSNscore. We found two bidirectional LSTM layers optimal for the spectrum channel, and three LSTM layers optimal for the light curve channels.}
    \label{fig:model}
\end{figure*}

We used \textit{Keras} \citep{chollet2015keras} Python library on top of TensorFlow \citep{tensorflow2015-whitepaper} framework for this study. We initially started with a single multiclass model to classify all CCSN subtypes similar to what is done in DASH \citep{Muthukrishna_2019}. During early training and validation, although the overall model accuracy was poor ($\sim$50\,\%), we recognized that the H-rich subtypes (Type II) formed a group, better separated from the H-poor subtypes (Type Ibc) group. Thus we decided to make an application divided into two hierarchical layers based on the current core-collapse SN classification scheme and use parallel binary classifiers (also known as the One vs. Rest strategy) in each layer instead of a single multiclass model. Our application's layer~1 has two binary classifiers, one trained for hydrogen-rich (H-rich) SNe and the second for hydrogen-poor (H-poor) SNe. Though this could have been a single binary classifier task, we train two parallel classifiers for an added layer of robustness. Layer~2a has three binary classifiers for three major subtypes of H-rich SNe (II -- normal Type II spectra, IIb-H -- IIb spectra with hydrogen present, IIn -- narrow lines from interaction) and layer~2b also has three models for major subtypes of H-poor SNe (Ib, Ic, Ic-BL). In total, we have 8 binary classification tasks. 

The number of samples in the training and test sets for each layer is outlined in Table~\ref{tab:numsamples}. The number inside brackets in the `training samples' column in Table~\ref{tab:numsamples} is the real number of training samples of that class, and the number outside brackets is the total training samples including the augmented data. The number of augmented samples added to each class is such that the total number of samples is roughly balanced across all classes in that layer, making the training samples of the sub-layers 2a and 2b not add up to the respective layer~1 numbers. The gold vs. bronze test set evaluation is only done for layer~1.

\begin{table}[h]
    \caption{Total number of samples in the training and test sets per binary classifier. The total number of training samples (including augmented data) per class is listed outside brackets in the second column, while the real number of training samples is listed inside brackets. The number of samples in the gold test set is listed outside brackets in the third column while the number of samples in the bronze set is listed inside brackets for Layer~1. The amount of augmented data added to each classifier in a layer is such that the samples per class are roughly balanced.}
    \label{tab:numsamples}
    \centering
    \begin{tabular}{>{\centering\arraybackslash}p{1.6cm}>{\centering\arraybackslash}m{2.8cm} >{\centering\arraybackslash}m{2.8cm}}
    \toprule
    \toprule
    Classifier & \# Training samples & \# Test samples \\
         & Total (real) & Gold (bronze) \\
     \midrule
    \multicolumn{3}{c}{Layer~1} \\
    \midrule
    H-rich & 8478 (6025)  & 627 (536)  \\
    H-poor & 8869 (2538)  & 153 (219) \\
    \midrule
    \multicolumn{3}{c}{Layer~2a} \\
    \midrule
    II & 6027 (4713)  & 911  \\
    IIb-H & 5500 (471) &  80  \\
    IIn & 5999 (841) &  172  \\
    \midrule
    \multicolumn{3}{c}{Layer~2b} \\
    \midrule
    Ib & 1761 (760)  & 110 \\
    Ic & 1775 (774) & 142 \\
    Ic-BL & 1844 (343) & 103 \\
    \bottomrule
    \end{tabular}
\end{table}

 The architecture of a single binary classifier model that we arrived at after the optimization process described in Section~\ref{sec:training} is shown in Figure~\ref{fig:model}. The models can be trained with up to five input channels, one for 1D optical spectra (the only required channel), two for 1D ZTF light curves (LC; $r$ and $g$ bands), and two for the $\delta m-\delta t$ representations of the 1D light curves. The multiple inputs are processed through separate network paths concatenated at the end, and the output is passed through a final dense layer with a sigmoid activation function to generate the final output probabilities. We experimented with several configurations by varying the kind of neural network layers in the channels and the number of layers per channel to arrive at the base architecture and then optimized its hyperparameters. 

 \section{Optimization and training}\label{sec:training}
 
 For the first step of the optimization process, we kept the number of NN layers constant (two per channel) and trained several models varying just the NN layer type. For spectra and 1D light curves, we decided to test 1D convolutional neural networks (CNN) and recurrent neural networks (RNN), specifically Gated Recurrent Unit (GRU), Long Short-Term Memory (LSTM) and bi-directional LSTM among the RNN layer types. For $\delta m-\delta t$ we decided to use 2D CNN layers. We used a 10\% dropout rate\footnote{Percentage of nodes intentionally dropped from the neural network to prevent overfitting} in between NN layers to tackle overfitting of the data. These initial models were compiled with the \texttt{Adam} optimizer (initial learning rate, $lr = 0.001$) and \texttt{BinaryCrossentropy} loss. The models were trained on H-rich and H-poor classification tasks with a batch size\footnote{The number of samples used in one training pass of the network} of 32, a validation split\footnote{Percentage of training data to be used for validation} of 33\%, and an early stopping criteria\footnote{A conditional criteria to stop the training of a model early if the loss does not decrease for a certain number (patience value) of epochs.} (patience value of 7 on validation loss). Validation accuracy and precision were used to decide the best model. We found that bi-LSTM layers performed best for the spectral channel and LSTM layers performed best for the 1D light curve channel, as they captured the connections present within these serial data. We then optimized the number of NN layers in each channel by varying them in one channel at a time between 1 to 4 NN layers, keeping the other channels and dropouts unchanged. The models were compiled, trained, and evaluated in the same manner as before and the optimal number of NN layers found per channel are shown in Figure~\ref{fig:model}. Next, with the help of \texttt{kerastuner}, we trained a grid of models varying the hyperparameter values within the ranges listed in Table~\ref{tab:tuner} to find the best-performing values. This tuning was done separately for H-rich and H-poor binary classifiers (all input channels used) and both classifiers settled to the same optimal hyperparameters. We applied further manual tuning to arrive at the final hyperparameter values shown in Figure~\ref{fig:model}. Our best-performing model favors high dropout rates similar to SNIascore. We found that the initial learning rate of the \texttt{Adam} optimizer, $lr=0.001$, and a mini-batch size of 64 performed well across all the binary classifiers.

   \begin{table}[h]
     \caption{Hyperparameter ranges for tuning binary classifiers with \texttt{kerastuner}.}
     \label{tab:tuner}
     \centering
     \begin{tabular}{>{\centering\arraybackslash}m{3cm}>{\centering\arraybackslash}m{2cm} >{\centering\arraybackslash}m{2cm}}
     \toprule
     \toprule
         Layers & Range & Step \\
          & (units) & (units) \\
    \midrule
        biLSTM  & 4 -- 24 & 4 \\
        LSTM  &  4 -- 24 & 4  \\
        Conv 2D (1)  &  16 -- 64 & 16  \\
        Conv 2D (2) & 8 -- 32 & 8 \\
        Dense  &  8 -- 64 & 8  \\
        Dropout &  0.1 -- 0.9 & 0.2  \\ 
    \bottomrule
     \end{tabular}
 \end{table}

 For the final training of the binary classifiers, we use a validation split of 0.33, 100 epochs, and apply early stopping criteria with a patience value of 7 on the validation loss metric. For balanced training of the binary classifiers in case one class has more samples than the other, we take all samples of the smaller class and choose an equal number of samples from the larger class for the first round of training. Then, we repeat the training by redrawing samples from the larger class without substitution until enough samples are left in the larger class or the training has been repeated 3 times. After all the binary classifiers have been trained and saved, we predict the final classifications on the test sets. We generate 100 predictions per sample with dropout enabled in the trained model (Monte Carlo Dropout technique) and calculate the mean and standard deviation of the 100 predicted probabilities to get the final predicted probability and the uncertainty on it. This is done for each sample to get predictions from each binary classifier. We use the following scheme to get the final classifications based on the probabilities given by the parallel classifiers. The classifier that provides the maximum probability is chosen as the final class (and the max probability as the final score) if the difference between the highest and the second highest probabilities is more than the sum of their uncertainties. The remaining samples are assigned an `ambiguous' classification and a score of zero.
 
 We also train sets of models for different input channel combinations (input cases) for each binary classification task to compare the contribution of the light curve input in different forms. The input cases are as follows: `only spectra', `spectra + 1D LC', `spectra + $\delta m-\delta t$', and `spectra + 1D LC + $\delta m-\delta t$' (all channels). The results of this comparison are presented in \S\ref{sec:results}.

\section{Performance}\label{sec:results}

\subsection{Layer~1: H-rich vs. H-poor}


\begin{figure*}[htbp!]
    \centering
    \subfloat[\small{Results from binary classifiers trained only using the spectral input channel.}]{%
    \includegraphics[width=0.9\textwidth]{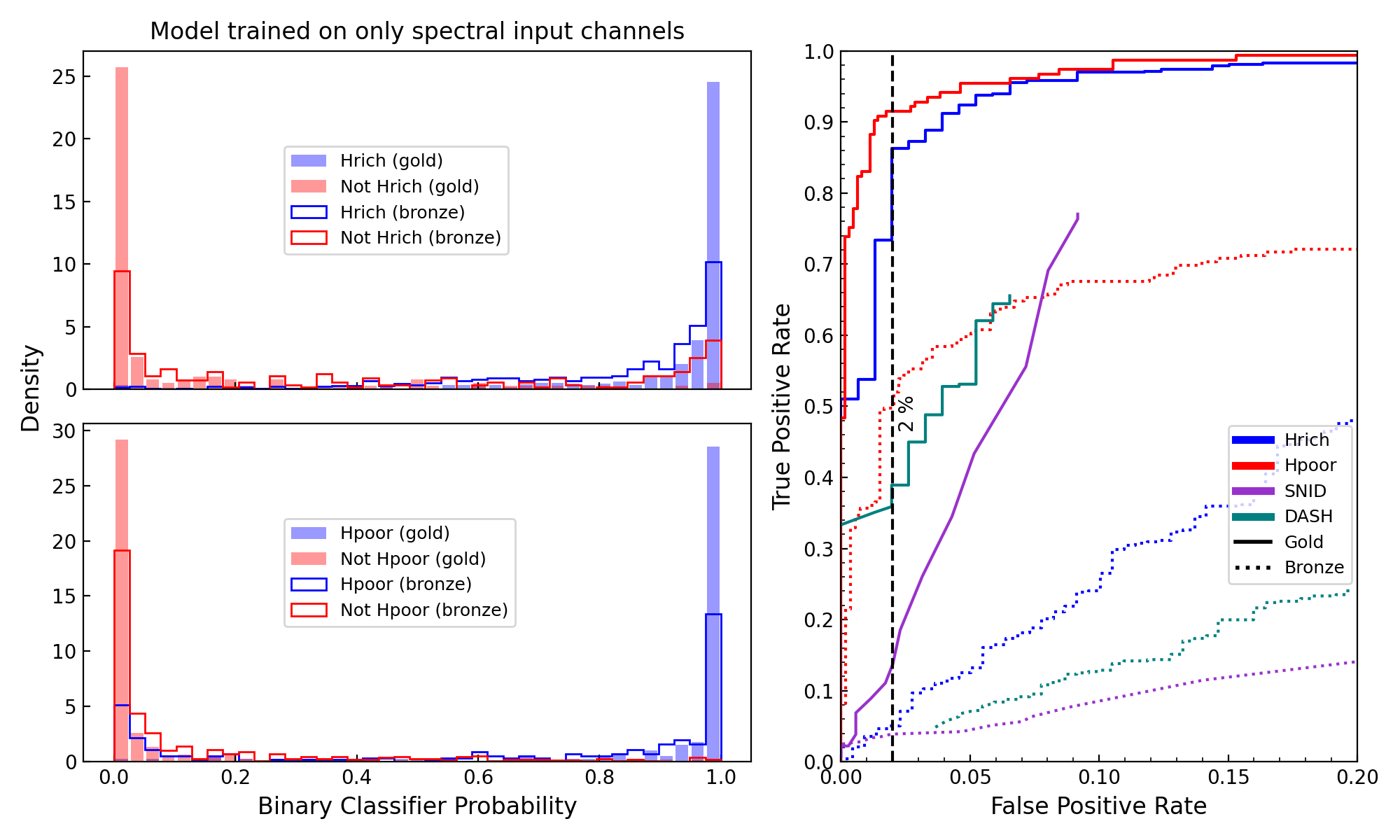}%
    \label{fig:histroc1a}}\\
    \subfloat[\small{Results from binary classifiers trained using all input channels, i.e.\ spectra, light curves, and $\delta m-\delta t$.}]{%
    \includegraphics[width=0.9\textwidth]{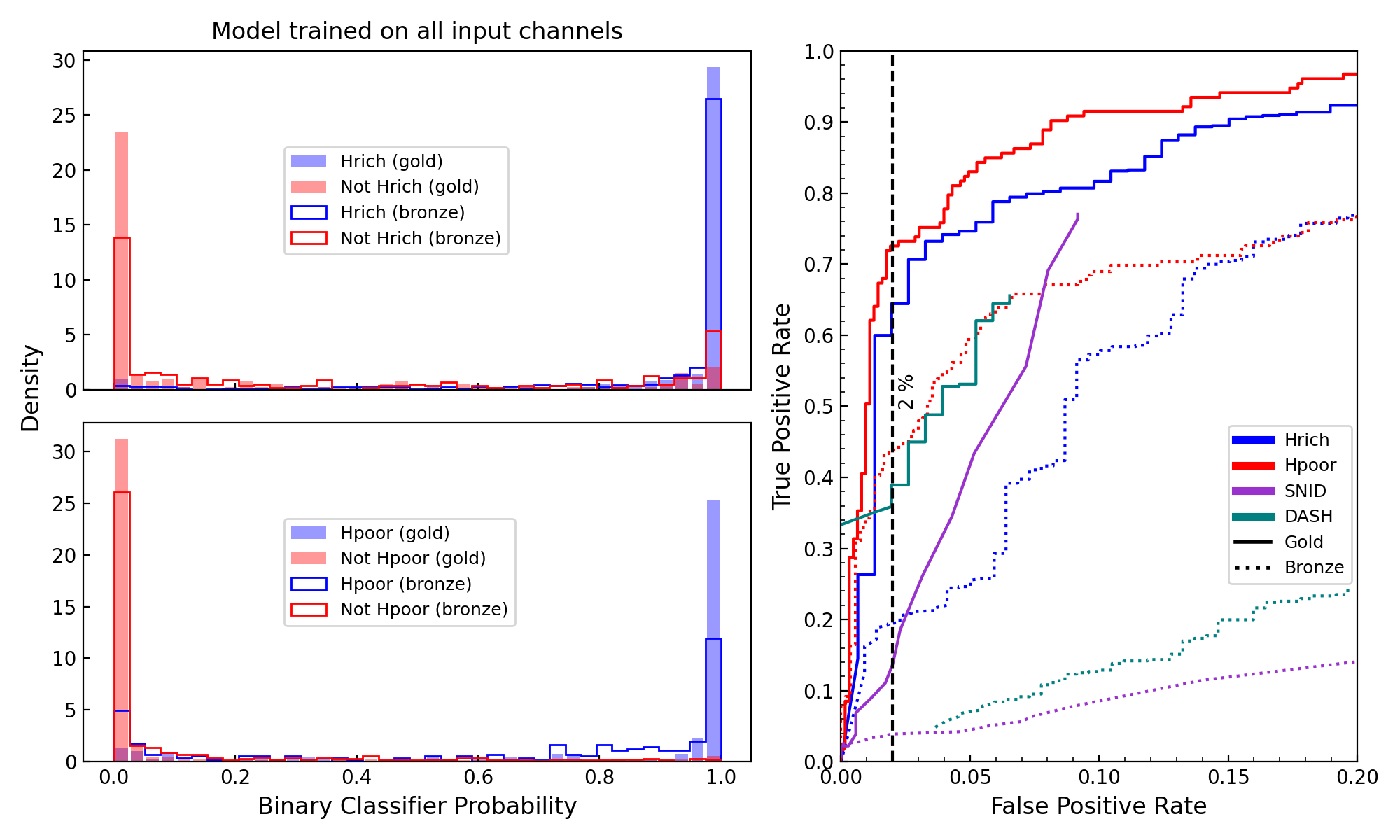}%
    \label{fig:histroc1b}}%
    \caption{Performance of layer~1 binary classifiers on gold and bronze test sets. \textit{Left} Distribution of binary classifier predicted probabilities for positive class (blue) vs. negative class (red) for H-rich and H-poor classifiers. The gold test set distribution is depicted as filled bars and the bronze set distribution as empty steps. \textit{Right} ROC curve from the two layer~1 binary classifiers (blue and red), with SNID (purple) and DASH (teal) for comparison. The gold set curves are depicted with solid lines and the bronze set curves with dotted lines. The black vertical dashed line marks a 2\% false positive rate.}
    \label{fig:histroc1}    
\end{figure*}

\begin{figure*}[htbp!]
    \centering
    \subfloat[\small{CMs constructed using all of the test samples' classifications. The `Ambi' column contains samples for which H-rich and H-poor classifiers assigned probabilities within their uncertainties. The overall classification accuracy for the gold set is highest in the `only spectra' case (without light curve addition), while for the bronze set is highest in the `spectra + 1D LCs' case. The utility of light curve input is reflected better in Figure~\ref{fig:conmats1b} when probability cuts are applied.}]{%
    \includegraphics[width=0.95\textwidth]{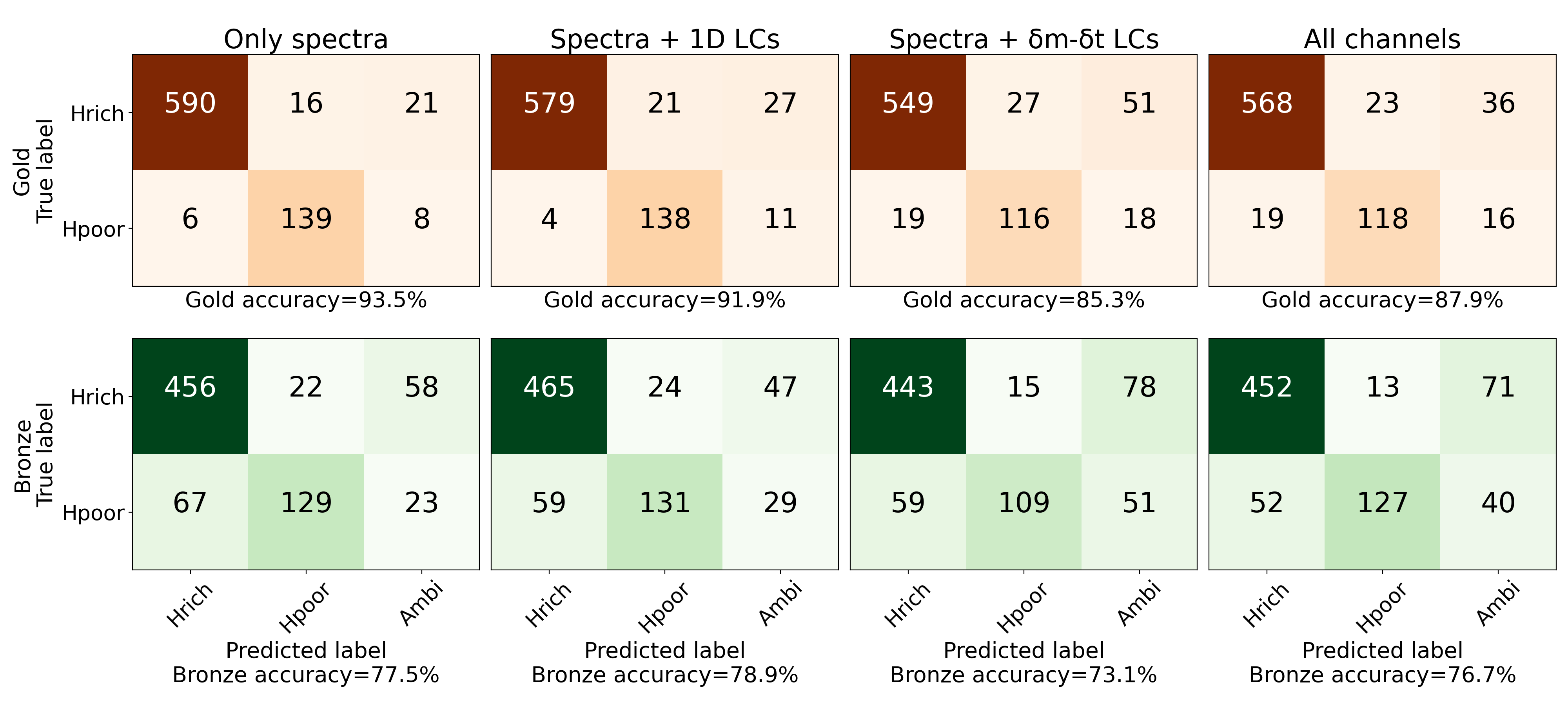}%
    \label{fig:conmats1a}}\\
    \subfloat[\small{CMs derived using only the test samples that pass confidence and uncertainty cuts. All the ambiguous cases get filtered out with a strict probability cut. The fraction of the test set that qualifies these cuts is printed under each confusion matrix. More bronze quality data gets reliable predictions when light curve inputs are used.}]{%
    \includegraphics[width=0.95\textwidth]{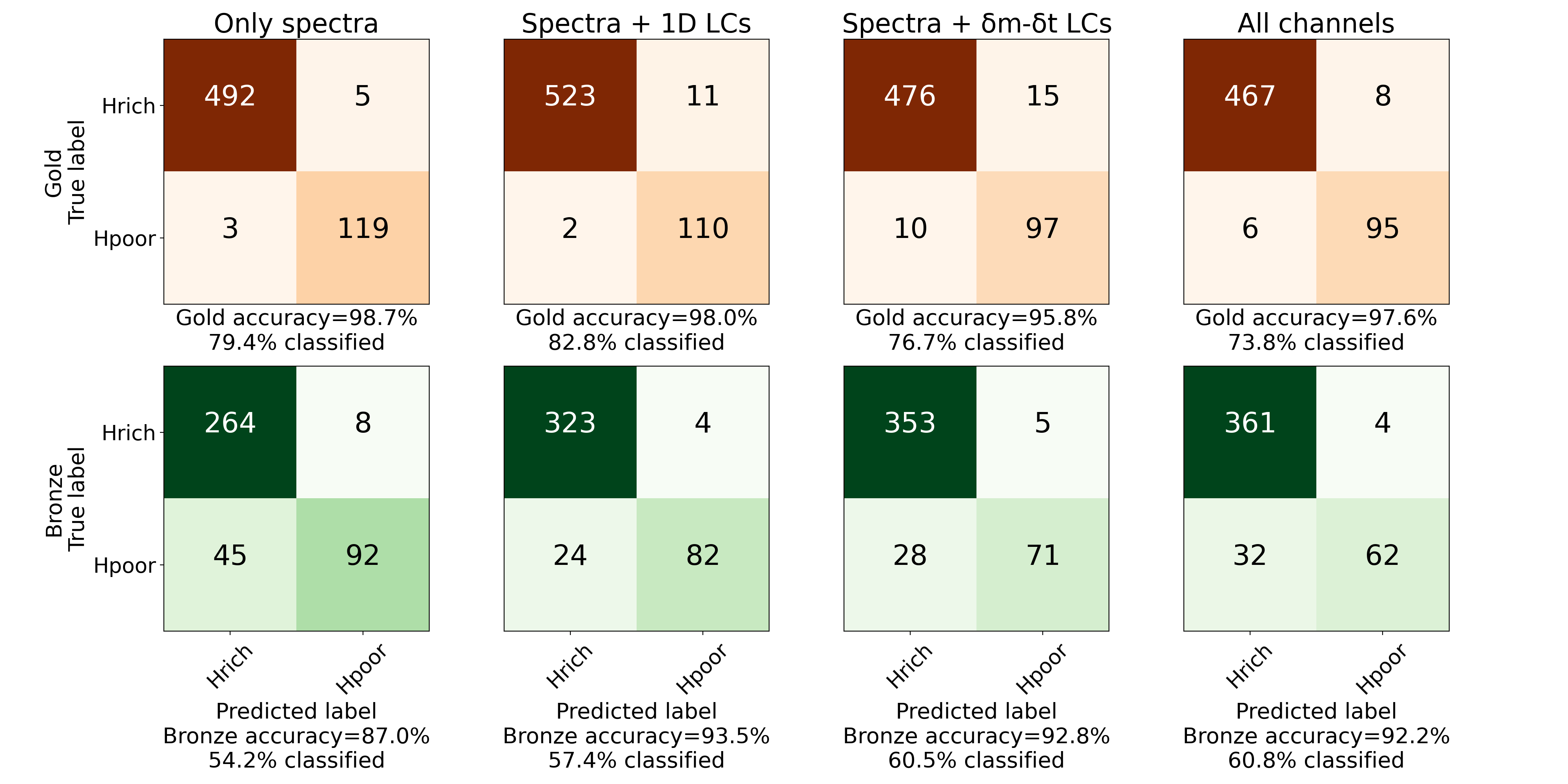}%
    \label{fig:conmats1b}}%
    \caption{Results from the layer~1 (H-rich vs. H-poor) models. The four columns of confusion matrices (CM) are for the four input cases --- `only spectra', `spectra + 1D LC', `spectra + $\delta m-\delta t$', and `spectra + 1D LC + $\delta m-\delta t$'. The top row of CMs in each subfigure is for the gold test set and the bottom row of CMs is for the bronze test set.}
    \label{fig:conmats1}
\end{figure*}

Figures~\ref{fig:histroc1} and \ref{fig:conmats1} show the performance of the layer~1 classification task i.e. H-rich vs. H-poor. Figure~\ref{fig:histroc1a} shows results for the ``only spectra" input case models and Figure~\ref{fig:histroc1b} for the ``all channels" input case models. The left panel in both subfigures shows the distribution of probabilities predicted for the gold (filled bars) and bronze (not filled bars) test sets, with blue bars denoting samples that belong to the class (positive) and red bars for samples that do not (negative). For any good binary classifier, this distribution should be highly bimodal with the positive class getting the highest probability scores and the negative class getting the lowest. Models of both input cases show such bimodal distributions, but the distribution is less sharp for the ``only spectra" input case than for the ``all channels" case. Thus, adding the light curve inputs to model training results in general higher predicted probabilities for the positive class and lower predicted probabilities for the negative class. However, it also increases the predicted probabilities of the false positive cases.

The right panel in both subfigures of Figure~\ref{fig:histroc1} shows the receiver operating characteristic (ROC) curve for the two layer~1 classifiers (in blue and red lines) on the gold and bronze test sets, along with the SNID (purple lines) and DASH (teal lines) ROC curves for comparison. A ROC curve depicts the true positive rate (TPR) vs. false positive rate (FPR) at all classification thresholds, which for our binary classifiers and DASH are the predicted probabilities while for SNID are the \texttt{rlap} scores (see Appendix~A for the method used to construct SNID ROC curves). The SNID and DASH ROC curves are calculated considering H-rich as the positive class. Our best classifiers achieve $\geq90$\% TPR compared to SNID's $\sim75$\% TPR at a FPR of 10\% and DASH's $\sim65$\% TPR at a FPR of $\sim6$\% on the gold set. Similar behavior is observed for the bronze set, with our best classifier achieving 30--60\% TPR versus SNID's and DASH's $\sim10$\% TPR at a FPR of 10\%. 

The addition of light curve inputs (both 1D LC and $\delta m-\delta t$ together) ends up reducing the performance of both classifiers on the gold set but improves the performance on the bronze set, especially for the H-rich case with its area-under-the-curve (AUC) being larger for the model trained on all channels (see solid and dashed lines in right panels of Figures~\ref{fig:histroc1b} and \ref{fig:histroc1a} respectively). This indicates that perhaps additional input is more suitable when the spectrum is of lesser quality, but can lead to more inaccuracies and model confusion otherwise. This could be due to how the multi-input channels are concatenated in CCSNscore architecture (see Figure~\ref{fig:model}), with the spectral channel getting less weight as more channels are added.

For directly sending classifications to TNS without human intervention, a model with extremely low FPR is preferred, similar to SNIascore \citep{Fremling_2021}. If 2\% FPR can be risked, $\sim80-90$\% of both H-rich and H-poor gold quality SNe can be sent to TNS using the best-performing models. For SNID this fraction is only $\sim10$\%. However, programs like SNID, as well as Superfit \citep{superfit} and Gelato \citep{gelato2008}), were always meant to assist with the classification process and thus perform best with manual user inputs, often providing more information than just the classification (for example, age, fairly accurate redshift, similarity to historical SNe, host galaxy characteristics, etc.). A more appropriate comparison can be made with DASH \citep{Muthukrishna_2019}, which also employs deep learning for automated classification purposes. In \citet{Muthukrishna_2019}, DASH's performance was tested on 212 spectra from OzDES ATELs released between 2015 and 2017 out of which 81\% were SNe Ia, and DASH provided correct classifications for 93\% of the 212 SNe. However, the same performance of DASH could not be attained on the SEDM spectra as tested in \citet{Fremling_2021}. \citet{kim_2024} compared the performance of SNID, NGSF \citep{ngsf}, and DASH on $\sim4600$ SEDM spectra and found the automatic accuracy for five-class classification task (Ia, II, Ibc, SLSN, notSN) to be $\sim63$\% (SNID), $\sim75$\% (NGSF) and $\sim62$\% (DASH). Particularly for CCSNe, DASH achieved only $\sim29$\% TPR at $\sim3$\% FPR for Type II and $\sim79$\% TPR at $\sim32$\% FPR for Type Ibc, which is not suitable for reporting classifications to TNS. We see the same with DASH's performance on our test sets, which is $\sim35$\% TPR for the gold set and $\sim4$\% TPR for the bronze set at a 2\% FPR.

Figure~\ref{fig:conmats1} displays the confusion matrices (CMs) for the gold and bronze test sets derived from the predictions of our layer~1 models (the four input cases). We present the CM data in `number of samples' instead of percentages to emphasize the number of misclassifications. The top row of CMs in both sub-figures (\ref{fig:conmats1a} \& \ref{fig:conmats1b}) are for the gold set and the bottom row of CMs are for the bronze set. The four columns of matrices are for the four input cases (from left to right, `only spectra', `spectra + 1D LC', `spectra + $\delta m-\delta t$', and `spectra + 1D LC + $\delta m-\delta t$'). The `accuracy' metric is printed under each CM. Henceforth, `only spectra', `spectra + 1D LC', `spectra + $\delta m-\delta t$', and `spectra + 1D LC + $\delta m-\delta t$' input cases will be referred to as cases `S', `SL', `SD', and `SLD' respectively. 

Figure~\ref{fig:conmats1a} shows the CMs constructed by including all samples of test sets without applying any quality cut on the predicted probabilities. As mentioned earlier in~\S\ref{sec:training}, the classifier that assigns the highest probability is chosen as the class for a test sample if the difference between the highest and the second highest probability is more than the sum of their uncertainties, otherwise, the test sample gets an `ambiguous' tag. These ambiguous cases occupy the third column in the CMs of Figure~\ref{fig:conmats1a}. When spectral quality is good (gold set), the case~S model performs better than other input cases, has the lowest number of false and ambiguous predictions, and the highest accuracy of 93.5\% (see Figure~\ref{fig:conmats1a}). When spectral quality is poor (bronze set), all cases perform similarly with case SL having slightly higher accuracy. The true usefulness of the light curve input can be seen in Figure~\ref{fig:conmats1b}, which shows the CMs constructed from the subset of the test sets filtered by cuts on predicted probability ($P$) and their uncertainties ($P_{unc}$) determined heuristically to obtain the most confident classifications. The threshold cuts are $P>0.8, P_{unc}<0.05$ for the gold set, and $P>0.9, P_{unc}<0.05$ for the bronze set (slightly stricter). The filtered subset fraction is printed under each matrix in Figure~\ref{fig:conmats1b}. A high probability threshold reduces false positives and increases accuracy, but discards the test samples that do not pass the threshold cuts. From Figure~\ref{fig:conmats1b}, we note that a higher fraction of bronze test samples pass the cuts in cases SL, SD, and SLD while maintaining high accuracies. For gold test samples, this holds for case~SL which has 82.8\% confident classifications compared to 79.4\% in case~S. This increase in high-confidence classifications also slightly increases false positives and negatives, which for the gold set reduces the accuracy in case~SL to 98\% from 98.7\% in case~S. But for the bronze set the overall accuracy in case~SLD still increases. Thus, multi-input channel models that can ingest auxiliary information relevant to classification are better suited for lesser-quality spectral data.

A caveat with our models is that they were trained on spectra that had been deredshifted (transformed from observed wavelengths to rest-frame wavelengths), and thus would require redshift information for real-time application. To analyze the effect of redshifting, we trained binary classifiers for H-rich and H-poor classes using only the spectral channel and redshifted spectra. We found that the classification accuracy for the whole set (gold $+$ bronze) is $\sim85.9$\%, the same as the performance of models trained on deredshifted spectra. This is expected as bi-directional LSTM layers are capable of capturing dependencies in sequences in both directions simultaneously. 

Another caveat with our models is that they were trained using the full duration of light curves but for real-time application, only epochs up until the spectral phase will be available (mostly early-time or pre-peak phase). Thus the real-time performance of the models using light curve input will be different than presented. Hence, we plan on using the Case~S models for real-time TNS reporting until the performance of partial light curve input is characterized.

Table~\ref{tab:results} further lists the following metrics that quantify the performance of our layer~1 models:

\begin{equation}
\begin{split}
 \mathrm{Accuracy} & = \frac{TP + TN}{TP + FN + FP + TN} \\
 \mathrm{Precision\ (or\ Purity)} & = \frac{TP}{TP + FP} \\ 
 \mathrm{TPR\ (or\ Recall)} & = \frac{TP}{TP + FN} \\
 \mathrm{FPR} & = \frac{FP}{FP + TN} \\
 \mathrm{F1score} & = \frac{2 \times \mathrm{Precision} \times \mathrm{Recall}}{\mathrm{Precision} + \mathrm{Recall}} 
\end{split}
\end{equation}

where $TP$ stands for true positives, $FP$ for false positives, $FN$ for false negatives and $TN$ for true negatives. Ambiguous cases were counted as $FN$ for the TPR calculation and $TN$ for the FPR calculation. 

\begin{table*}[htbp!]
    \caption{Performance metrics of the layer~1 task models on the gold and bronze test sets. The total accuracy in the final row shows the combined performance on the gold and bronze sets.}
    \label{tab:results}
    \centering
    \begin{tabular}{>{\centering\arraybackslash}m{1cm} >{\centering\arraybackslash}m{1.5cm} >{\centering\arraybackslash}m{1.2cm} >{\centering\arraybackslash}m{1.2cm} >{\centering\arraybackslash}m{1.2cm} >{\centering\arraybackslash}m{1.2cm} >{\centering\arraybackslash}m{1.2cm} >{\centering\arraybackslash}m{1.2cm} >{\centering\arraybackslash}m{1.2cm} >{\centering\arraybackslash}m{1.2cm}}
    \toprule 
    \toprule 
    \multicolumn{2}{c}{} & \multicolumn{2}{c}{Only Spectra} & \multicolumn{2}{c}{Spectra + 1D LCs} &
     \multicolumn{2}{c}{Spectra + $\delta m-\delta t$} & \multicolumn{2}{c}{All channels} \\
    \multicolumn{2}{c}{} & \multicolumn{2}{c}{Case S} & \multicolumn{2}{c}{Case SL} & \multicolumn{2}{c}{Case SD} & \multicolumn{2}{c}{Case SLD} \\
     \midrule
    & & H-rich & H-poor & H-rich & H-poor & H-rich & H-poor & H-rich & H-poor \\
     \midrule
    & F1 score & 96.5\% & 90.3\% & 95.7\% & 88.5\% & 91.9\% & 78.4\% & 93.6\% & 80.3\% \\
    \multirow{3}{1cm}{Gold} &  Precision  & 99.0\% & 89.7\% & 99.3\% & 86.8\% & 96.7\% & 81.1\% & 96.8\% & 83.7\% \\
     & TPR  & 94.1\% & 90.8\% & 92.3\% & 90.2\% & 87.6\% & 75.8\% & 90.6\% & 77.1\% \\
     &  FPR  & 3.9\% & 2.6\% & 2.6\% & 3.3\% & 12.4\% & 4.3\% & 12.4\% & 3.7\%  \\
     & Accuracy & \multicolumn{2}{c}{93.5\%} & \multicolumn{2}{c}{91.9\%} & \multicolumn{2}{c}{85.3\%} & \multicolumn{2}{c}{87.9\%} \\ 
     \midrule
    & F1 score & 86.1\% & 69.7\% & 87.7\% & 70.1\% & 85.4\% & 63.6\% & 86.9\% & 70.8\% \\
    \multirow{3}{1cm}{Bronze} &   Precision  & 87.2\% & 85.4\% & 88.7\% & 84.5\% & 88.2\% & 87.9\% & 89.7\% & 90.7\% \\
     &   TPR  & 85.1\% & 58.9\% & 86.8\% & 59.8\% & 82.6\% & 49.8\% & 84.3\% & 58.0\% \\
     &  FPR  & 30.6\% & 4.1\% & 26.9\% & 4.5\% & 26.9\% & 2.8\% & 23.7\% & 2.4\% \\   
     & Accuracy & \multicolumn{2}{c}{77.5\%} & \multicolumn{2}{c}{78.9\%} & \multicolumn{2}{c}{73.1\%} & \multicolumn{2}{c}{76.7\%} \\ 
     \midrule
     Total & Accuracy & \multicolumn{2}{c}{85.6\%} & \multicolumn{2}{c}{85.5\%} & \multicolumn{2}{c}{79.3\%} & \multicolumn{2}{c}{82.4\%} \\ 
     \bottomrule
    \end{tabular}
\end{table*}

\subsubsection{Layer~1 misclassifications}
 From Figure~\ref{fig:conmats1b}, there are 3, 2, 10, and 6 samples of the H-poor class from the gold set that get misclassified as H-rich even with confidence cuts in cases S, SL, SD, and SLD respectively. There are 14 unique misclassified samples out of these 21, with some common samples among cases, and the misclassifications share some similarities. The misclassifications in cases~S and SL have weak H-poor spectral line features with a strong blue continuum, making them look similar to early-time H-rich spectra. Out of the 10 misclassifications in case~SD, 2 are the same as case~S, and the remaining 8 have either long declining or peculiar light curves (unlike regular Type~Ibc SNe) possibly influencing the incorrect decision. Finally, of the 6 case~SLD misclassifications, 3 are common with case~SD, and the remaining 3 are cases that completely lack $g$-band coverage.

 Similarly, 5, 11, 15, and 8 gold samples of the H-rich class are misclassified as H-poor in cases S, SL, SD, and SLD respectively. Out of the 5 case~S misclassifications, 3 are nebular spectra, 1 does not show H$\alpha$ emission and 1 seems to be a genuine mistake. Out of the 11 case~SL misclassifications, 7 are nebular spectra from the same supernova, SN~2023rky -- a Type IIL (that have a similar light curve shape to H-poor SNe), 1 does not show strong H$\alpha$ emission, and the remaining 3 are genuine mistakes. Nine of 15 Case~SD misclassifications are also from SN~2023rky, 2 do not show H$\alpha$ emission and 4 are genuine mistakes. For case~SLD, 6 out of 8 misclassifications are from SN~2023rky and the remaining 2 are genuine mistakes. There are 20 unique misclassified samples across all cases (39 total) from 12 unique SNe, and 9 samples belong to just one SN.

 Considering the bronze set, there are many H-poor SNe misclassified as H-rich (Figure~\ref{fig:conmats1a}), most of which are because of host contamination. For example, 56 out of the 67 bronze set misclassifications in case~S are due to host contamination. The host-contamination cases are inherently difficult to classify with SEDM spectra even with the help of SNID or NGSF, and thus often require intervention at the level of raw data reduction. Another observation can be made from Figure~\ref{fig:conmats1a} for the ambiguous cases. The bronze set has more ambiguous classifications for all the input cases, with more H-rich SNe classified as ambiguous (likely due to a higher occurrence of blue featureless spectra).

\subsection{Layer~2: Sub-typing of H-rich and H-poor SNe}

\begin{figure*}
    \centering
    \subfloat[\small{Confusion matrices from the layer~2a (II, IIb-H, IIn) models.}]{%
    \includegraphics[width=0.95\textwidth]{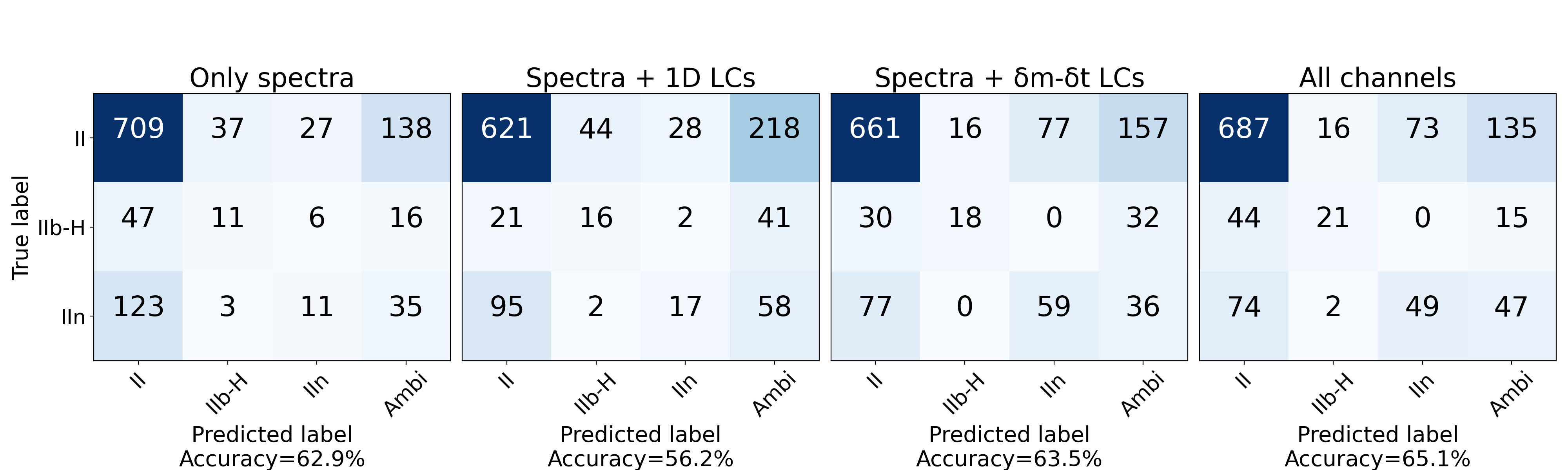}%
    \label{fig:conmats2a}}\\
    \subfloat[{Confusion matrices from the layer~2b (Ib, Ic, Ic-BL) models.}]{%
    \includegraphics[width=0.95\textwidth]{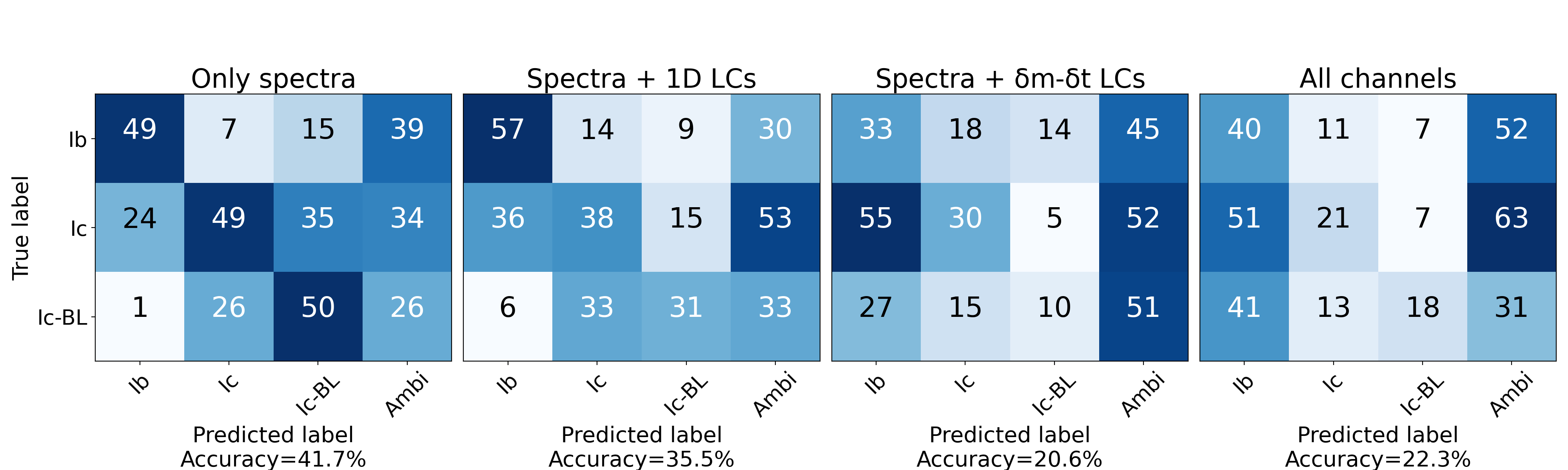}%
    \label{fig:conmats2b}}
    \caption{Confusion matrices for the whole test set derived from the layer~2 models. The four columns in subfigures are for the four input cases --- `only spectra', `spectra + 1D LC', `spectra + $\delta m-\delta t$', and `spectra + 1D LC + $\delta m-\delta t$'. These confusion matrices were constructed considering all of the test samples' classifications.}
    \label{fig:conmats2}
\end{figure*}

The performance of CCSNscore's layer~2a and layer~2b models which are trained for classification into subtypes of Type~II and Type~Ibc respectively are presented in Figure~\ref{fig:conmats2}. The confusion matrices are created with the full set instead of splitting into gold and bronze sets as the rarer subsets already have limited samples. Looking at the effect of light curve input on classification accuracy, layer~2a (SN II subtypes; Figure~\ref{fig:conmats2a}) seems to benefit marginally from the light curve input (particularly $\delta m-\delta t$). While cases~SL, SD, and SLD have fewer Type~IIb-H and Type~IIn samples misclassified as normal Type~II, they also have more normal Type~II misclassified as the `IIn' or ambiguous.

The opposite effect of light curve addition is seen for layer~2b (SN Ibc subtypes; Figure~\ref{fig:conmats2b}). Cases SL, SD, and SLD have lower accuracies than in case~S, with case~SD performing the worst. As SESN light curve properties do not differ much among the subtypes, perhaps the extra input information lends to confusion in the models. On the other hand, SN~II light curve properties of normal Type II, IIb, and IIn at least show some variety, thus making the extra input marginally useful. From Figure~\ref{fig:conmats2b}, the case~S accuracy is the highest at 41.7\%, but many misclassifications are among Ic and Ic-BL samples. If Ic and Ic-BL samples are considered just one class (Ic) and accuracy is measured for a Type~Ib vs. Type~Ic classification, case~S accuracy gets bumped up to 58.7\% including ambiguous samples. If ambiguous samples are not considered, 72\% of the total samples get a non-ambiguous (Ib or Ic) classification out of which 81.7\% are correct. Though this might not be robust enough for fully automated classification, the subtype predictions from CCSNscore can be provided as additional information with the TNS reports.

\section{Real-time implementation}

CCSNscore will be integrated into the current SEDM pipeline which already runs SNID and SNIascore. It will be used in conjunction with SNIascore to infer the classification for real-time application. The following scenarios are possible based on figure~4 of \citet{Fremling_2021}. First, $\sim85$\% of true SNe~Ia and $\sim0.5$\% of true CCSNe are likely to pass $SNIascore>0.6$ cut, which has generally been robust enough for automated SN~Ia classification reporting. Since CCSNscore has not been trained on SN~Ia data, its performance on real SNe~Ia is unreliable, and thus using CCSNscore prediction in this scenario will not help recover the 0.5\% true CCSNe. Therefore, CCSNscore predictions will not be considered where $SNIascore>0.6$. Next, $\sim12$\% true SNe~Ia and $\sim9.5$\% true CCSNe are likely to get $0.1 < SNIascore < 0.6$. In this scenario, a H-rich classification by CCSNscore can mean host-contamination in the spectrum (which should not be sent to TNS), a bad classification, or a true SN~II. A H-poor classification by CCSNscore does not provide any distinguishing information as SNe~Ia are also hydrogen-poor. We will flag these cases for visual inspection and not report them automatically. In total, $\sim10$\% of true CCSNe will not meet automatic classification criteria during real-time operations. Finally, $\sim90$\% true CCSNe and $\sim3$\% true SNe~Ia will likely pass the $SNIascore<0.1$ cut. The 3\% SN~Ia false negatives will comprise difficult cases of host contamination, peculiar SNe~Ia, or bad classifications. From preliminary analysis, we found that with the strict threshold cuts on $CCSNscore$ and $numSNID$ that we apply for TNS reporting (described below), most of these SN~Ia false negatives get filtered out and the number of true SNe~Ia misclassified as CCSNe that might get sent to TNS goes down to $\sim0.5$\%. From the BTS survey, $\sim800$ SNe with $m_{pk}\leq18.5$ are expected to be classified as SNe~Ia annually, which translates to $\sim 24$ SNe~Ia getting a $SNIascore<0.1$ and only $\sim4$ SNe~Ia per year passing the further TNS reporting criteria, which is a small number objectively. Therefore, we will use this criteria ($SNIascore<0.1$) to filter out potential CCSNe from SNe~Ia and $CCSNscore$, and $numSNID$ quality criteria will be used to determine their automatic reporting eligibility. Note that these numbers may change over time based on the real-time performance of the classifiers, so we will adjust the thresholds later on if needed\footnote{The current TNS reporting criteria cuts are posted on the \href{https://github.com/Yashvi-Sharma/CCSNscore/tree/main/data}{CCSNscore GitHub page} and will be updated there if any changes are made.}.

As the performance of models that include light curve input on only early-time light curve data has not been characterized, we will use the layer~1 `only spectra' model predictions for TNS reporting for now while we further analyze the light curve input performance. Therefore, the $CCSNscore$ and $CCSNscore_{unc}$ are the prediction probability and uncertainty on prediction probability, respectively, obtained from the `only spectra' model for H-rich (Type~II) vs. H-poor (Type~Ibc) task. The TNS reporting criteria will be as follows:

\begin{enumerate}
    \item For $numSNID \geq 30$: $CCSNscore > 0.8$ and $CCSNscore_{unc} < 0.05$
    \item For $20 \leq numSNID < 30$: $CCSNscore > 0.9$ and $CCSNScore_{unc} < 0.05$
    \item For $numSNID < 20$: $CCSNscore > 0.95$ and $CCSNscore_{unc} < 0.05$
\end{enumerate}

The number of samples that pass the above three criteria in our full test set (1535 samples) are 634, 49, and 275, respectively ($62.4\%$ of the full test set). Out of these, 598, 48, and 255 are correct classifications, which corresponds to a $\sim94\%$ accuracy rate and 6\% misclassifications. Among the SESNe in the full test set, 174, 13 and 59 pass the three TNS reporting criteria respectively, which corresponds to $\sim66\%$ of the total SESNe (372). Again from the BTS survey, $\sim81$ SNe with $m_{pk}\leq18.5$ are expected to be classified as SESNe annually, and therefore $\sim66\%$ of them or $\sim53$ will pass the TNS reporting criteria. Therefore, the $\sim4$ SNe~Ia that are expected to be falsely reported to TNS as SESNe annually will make up $\sim7.5\%$ of the total reported SESNe classifications. Improving the SN~Ia automatic classifiers can help reduce these false positives in the future.

A breakdown of the true positive rate by phase and true type is shown in Table~\ref{tab:real} for the samples that pass the TNS criteria. We do not see a significant difference with phase for Type~II samples under any criteria. However, Type~Ibc samples under $numSNID \geq 30$ have a higher true positive rate and precision for earlier phases (phase$\leq10$ days), while for $numSNID < 20$, the true positive rate and precision are higher for the later phases. This indicates that the difficult cases have a better chance of successful classification with post-peak spectra. We plan on further testing and revising this selection scheme over the next few months before actually starting real-time TNS reporting.

\begin{table}[htbp!]
    \caption{Performance of the layer~1 `only spectra' model predictions on the test samples that pass the TNS reporting criteria. The true positive rate and precision values are reported separately by phase ($\leq10$ days or $>10$ days) and the true type (II and Ibc). The number of samples that pass the different TNS criteria is shown in brackets in the first column.}
    \label{tab:real}
    \centering
    \begin{tabular}{>{\centering\arraybackslash}m{0.8cm} >{\centering\arraybackslash}m{1.3cm} | >{\centering\arraybackslash}m{1cm} >{\centering\arraybackslash}m{1cm} | >{\centering\arraybackslash}m{1cm} >{\centering\arraybackslash}m{1cm}}
    \toprule
    \toprule
       \multicolumn{2}{c}{True class:}  & \multicolumn{2}{c}{Type~II} & \multicolumn{2}{c}{Type~Ibc} \\
    \midrule
    \multicolumn{2}{c}{Phase:} & $\leq10d$ & $>10d$ & $\leq10d$ & $>10d$ \\
     \midrule
     \multirow{2}{0.8cm}{1. (634)} & TPR & 99.7\%  & 97.5\% & 83.2\% & 80.3\% \\
     & Precision & 94.0\% & 92.8\% & 98.9\% & 92.4\% \\
     \midrule
     \multirow{2}{0.8cm}{2. (49)} & TPR & 100\% & 100\% & 87.5\% & 100\% \\
     & Precision & 96.4\% & 100\% & 100\% & 100\% \\
     \midrule
     \multirow{2}{0.8cm}{3. (275)} & TPR & 96.4\% & 95.7\% & 78.4\% & 81.8\% \\
     & Precision & 95.3\% & 91.8\% & 82.8\% & 90.0\% \\
     \bottomrule
    \end{tabular}
\end{table}

\section{Discussion and Summary}\label{sec:disc}
In this work, we have presented a new deep-learning based software for the automated classification of core-collapse supernovae from their ultra-low resolution spectra (from SEDM) called CCSNscore. CCSNscore consists of hierarchical classification tasks, with its layer~1 meant for classification between Type II (or H-rich) and Type Ibc (or H-poor) SNe, layer~2a meant for classification between subtypes of Type II (II, IIb, IIn), and layer~2b meant for classification between subtypes of Type Ibc (Ib, Ic, Ic-BL). CCSNscore can be trained with up to five input channels, the spectral channel for SEDM spectra being the required input, and the ZTF $g$- and $r$-band 1D light curves and their respective $\delta m-\delta t$ representations as four additional inputs. We trained four different models for four input cases (`only spectra', `spectra + 1D LC', `spectra + $\delta m-\delta t$', and `spectra + 1D LC + $\delta m-\delta t$') and quantified the benefit of light curves to the classification process. We list our main results below:

\begin{itemize}
    \item CCSNscore's layer~1 performance with just spectral data input is quite robust for real-time TNS reporting of the classifications when strict score and uncertainty cuts are applied. Adding light curve input boosts the number of samples with high scores while maintaining accuracy, which can be useful for difficult classification cases. With just spectra, only 79.4\% (54.2\%) of the gold (bronze) test set pass the threshold cuts, out of which 98.7\% (87\%) are accurate. Comparatively, with spectra and 1D light curves, 82.8\% (57.4\%) of the gold (bronze) test set pass the threshold cuts, out of which 98\% (93.5\%) are accurate. 
    \item CCSNscore also provides subtype predictions from its layer~2 models, which can assist astronomers with manual classifications. The light curve input marginally improves accuracy over the `only spectra' case in the Type II subtyping task. On the other hand, given that SN Ib and Ic light curves are very similar, the light curve input actually reduces the classification accuracy for the Type Ibc subtyping task.
    \item RNN architecture seems optimal for ultra-low resolution spectral sequences, as found in this study and in \citet{Fremling_2021}. The biLSTM layers are also capable of deducing classification without needing redshift correction. 
    \item CCSNscore misclassifies H-poor SNe as H-rich more frequently than it does the other way around because of host-contamination cases, thus making H-poor predictions more reliable until host-contamination is dealt with at the data processing level. The misclassified true H-rich samples come from multiple spectra of a few unique SNe. The misclassified true H-poor samples come less often from multiple spectra of a few unique SNe but often share similar characteristics (weak features, strong host lines, etc.).
    \item We will use SNIascore to filter out likely SNe~Ia and select potential CCSNe candidates by applying a threshold cut of $SNIascore<0.1$. Then, we will apply threshold cuts on $numSNID$, $CCSNscore$, and $CCSNscore_{unc}$ to determine eligible candidates for reporting to TNS as described in \S\ref{sec:realtime}. Based on the BTS statistics on bright transients ($m_{pk}\leq18.5$) and expected SNIascore and CCSNscore performance, we expect $\sim0.5\%$ ($\sim4$) of true SNe~Ia to be misclassified as SNe~Ibc on TNS annually. For the classification of CCSNe into Type~II or Type~Ibc, we expect $\sim62\%$ of the total real-time true CCSNe spectra to qualify for TNS reporting, out of which we expect $\sim94\%$ correct classifications and $\sim6\%$ misclassifications.
\end{itemize} 

\citet{kim_2024} suggest that the effect of instrument resolution might not be significant for SEDM spectra when it comes to classification by SNID, NGSF, and DASH as all three programs appropriately preprocess the input (including smoothing and binning) to compare to the templates. But still, DASH's performance on SEDM spectra does not compare to models trained specifically using SEDM spectra like SNIascore and CCSNscore. This arises from the fact that deep learning models can be sensitive to the data they are trained with and do not generalize well until trained with an extremely large quantity of data. \citet{kim_2024} also note that a two-category classification task has more accurate results than a five-category task as CCSNe are more difficult to classify than SNe~Ia, and CCSNe need more than just spectral information to be classified robustly. CCSNscore addresses the above issues by 
\begin{enumerate}
    \item increasing the training data by including spectra from the Open Supernova Catalog smoothed to match the varying resolutions to the SEDM training set,
    \item splitting the classification tasks into hierarchical layers which are based on the traditional supernova classification scheme and training parallel binary classifiers instead of a single multi-class classifier,
    \item using a model architecture that can ingest multiple types of inputs, and using photometry data as an additional input.
\end{enumerate}

The dominant source of CCSNscore's misclassifications is host contamination, which is difficult to eliminate from the final processed spectra. SEDM's resolution augments the issue by blending the strong host lines around H$\alpha$, making them appear like SN II or SN IIn H$\alpha$ features. There are efforts to separate the SN light from its host galaxy at the data reduction level for SEDM through careful contour separation \citep{Kim2022} and hyperspectral scene modeling of the host galaxy \citep{Lezmy2022}, which could greatly improve CCSNscore's automatic typing accuracy.

Another possible method to improve the fidelity of CCSNscore could be changing the way photometric input is supplied and modeled. A well-performing photometric supernova classifier architecture from the literature that has already been tested independently \citep{Charnock2017,Pelican2019,Moller2020,Burhanudin2022,Allam2023} could be added as an additional input channel. Moreover, models that can predict the classification using only early-time photometric data and host-galaxy information will be the most useful as additional channels to CCSNscore \citep{Rapid2019,Scone2022}.

CCSNscore is a small step towards handling the increased load of transient spectroscopic data, which will become more important for future photometric and spectroscopic surveys. Currently, CCSNscore is only capable of providing a robust broad Type (II vs. Ibc) and a prediction for a subtype (less accurately) but does not provide other crucial information that can be deduced from spectra (redshift and phase). Future work for CCSNscore could focus on attempting redshift prediction using a similar model structure and exploring auxiliary inputs that can aid in such a task. 


\section{Acknowledgment}

 Based on observations obtained with the Samuel Oschin Telescope 48-inch and the 60-inch Telescope at the Palomar Observatory as part of the Zwicky Transient Facility project. ZTF is supported by the National Science Foundation under Grants No. AST-1440341 and AST-2034437 and a collaboration including current partners Caltech, IPAC, the Weizmann Institute of Science, the Oskar Klein Center at Stockholm University, the University of Maryland, Deutsches Elektronen-Synchrotron and Humboldt University, the TANGO Consortium of Taiwan, the University of Wisconsin at Milwaukee, Trinity College Dublin, Lawrence Livermore National Laboratories, IN2P3, University of Warwick, Ruhr University Bochum, Northwestern University and former partners the University of Washington, Los Alamos National Laboratories, and Lawrence Berkeley National Laboratories. Operations are conducted by COO, IPAC, and UW. The ZTF forced-photometry service was funded under the Heising-Simons Foundation grant \#12540303 (PI: Graham). The SED Machine is based upon work supported by the National Science Foundation under Grant No. 1106171. This work was supported by the GROWTH project \citep{Kasliwal_2019} funded by the National Science Foundation under Grant No 
1545949. MMK acknowledges generous support from the David and Lucille Packard Foundation. The Gordon and Betty Moore Foundation, through both the Data-Driven Investigator Program and a dedicated grant, provided critical funding for SkyPortal. 

Y. Sharma thanks the LSSTC Data Science Fellowship Program, which is funded by LSSTC, NSF Cybertraining Grant \#1829740, the Brinson Foundation, and the Moore Foundation; her participation in the program has benefited this work. M.W.C acknowledges support from the National Science Foundation with grant numbers PHY-2308862 and PHY-2117997. Zwicky Transient Facility access for N.R. was supported by Northwestern University and the Center for Interdisciplinary Exploration and Research in Astrophysics (CIERA). Y.-L.K. has received funding from the Science and Technology Facilities Council [grant number ST/V000713/1]. 

\facilities{Palomar 60-in. (SEDM)}

\software{pandas \citep{pandas}, NumPy \citep{numpy}, astropy \citep{astropy2013,astropy2018}, Keras \citep{chollet2015keras}, TensorFlow \citep{tensorflow2015-whitepaper}}

\appendix
\section{SNID ROC curve}
To construct the SNID ROC curve of Figure~\ref{fig:histroc1}, we ran SNID on our gold and bronze test sets with the following settings. We set $lapmin=0.1$, $rlapmin=0$, set `Ia' and `notSN' as template types to avoid (as we wanted to obtain only CCSN predictions), set maximum redshift $zmax=0.2$, and ran SNID with interactive and plotting disabled on all spectra. The top SNID match of each sample was set as the automatically predicted classification from SNID. For the subset of spectra that did not get any template matches with the above settings, we gradually increased $zmax$ to 0.5 until all samples had at least one match. The predicted classifications were then labeled as H-rich or H-poor based on the SNID assigned type, and the $rlap$ scores were used to measure the confidence in those predictions. Then to calculate the ROC curve, we used $rlap$ thresholds in the range 0--25 with a step of 1, and at each threshold, we calculated the true positive rate and the false positive rate, considering H-rich as the true positive class.

\bibliography{biblio}
\bibliographystyle{aasjournal}

\end{document}